\documentclass[aps,twocolumn,pra]{revtex4}
\usepackage{amsmath}
\usepackage{amsfonts}
\usepackage{amssymb}
\usepackage{epsfig}
\usepackage{graphicx}%floatfix,superscriptaddress
\begin{document}
\title{Double Electromagnetically Induced Transparency and Narrowing of Probe Absorption in a Ring Cavity with Nanomechanical Mirrors}
\author{Sumei Huang$^{1,2}$}
\address{$^1$ Department of Physics, University of California, Merced, California 95343, USA\\
$^2$ Department of Electrical and Computer Engineering,
National University of Singapore, 4 Engineering Drive 3, Singapore 117583}

\date{\today}
\begin{abstract}
We study the effect of a strong coupling field on the absorptive property of a ring cavity with two mirrors oscillating at slightly different frequencies to a weak probe field. We observe double electromagnetically induced transparency windows separated by an absorption peak at line center in the output probe field under the action of a strong coupling field. We find that increasing driving power can broaden the two transparency windows, which results in narrowing of the central absorption peak. At high driving power, the linewidth of the sharp central absorption peak is approximately equal to the mechanical linewidth. We show the normal mode splitting in both the output probe field and the generated Stokes field. We also find that the suppression of the four-wave mixing process can be achieved on resonance.
\end{abstract}
\maketitle
\section{Introduction}

It is well-known that a $\Lambda$-type three-level atomic medium can become transparent to a weak probe field by applying a strong coupling field, which is the result of the destructive interference between two different excitation pathways to the upper level. This is the phenomenon of electromagnetically induced transparency (EIT) \cite{EIT1, EIT2, EIT3, EIT4}. The EIT has been shown to be important for various applications such as slow light \cite{Hau}, light storage \cite{Liu}, and so on. Besides, the studies of EIT have been extended to multi-level atomic systems. The double EIT windows separated by a narrow absorption peak in the probe absorption spectrum have been observed in the four-level atomic systems \cite{Lukin,Gavra,Goren,Scully,Knight}. Recently, the EIT effect has been reported in the macroscopic optomechanical systems. The analogy of EIT in optomechanical systems has been shown theoretically \cite{Agarwal1}, and observed in a number of experiments in optical cavities \cite{Weis, Lin, Safavi} and microwave cavities \cite{Teufel,Ncomms}. Moreover EIT in optomechanical systems in the nonlinear regime was analyzed \cite{Lemonde,Girvin,Kronwald}. Additionally, the electromagnetically induced absorption, the opposite effect to EIT, was discussed in a two-cavity optomechanical system \cite{Qu2}. In addition, it has been proven that a strong dispersive coupling between the optical mode and the mechanical mode when the effective optomechanical coupling rate exceeds the optical and mechanical decay rates leads to normal mode splitting \cite{Florian,Kippenberg,Aspelmeyer}. The effective optomechanical coupling rate can be enhanced by increasing the power of the driving laser.

In this paper, we investigate the nonlinear response of a ring cavity with two moving mirrors having two close frequencies to a weak probe field in the absence and the presence of a strong coupling field. We find that there are two transparency windows and an absorption peak in the transmitted probe field in the presence of the coupling field. The pump-induced broadening of the two EIT dips leads to narrowing the central absorption peak. And the narrow absorption peak associated with double EIT windows may have potential application in high-resolution laser spectroscopy \cite{Lukin,Gavra,Goren}. We also observe the normal mode splitting in the output probe field at high driving power. In addition, we show that the normal mode splitting occurs in the Stokes field generated by means of the four-wave mixing process. And the four-wave mixing process is completely suppressed on resonance. However, if two movable mirrors in a ring cavity are oscillating at identical frequencies, the EIT-like dip can be observed in the output probe field, and the four-wave mixing process is not suppressed on resonance.

The paper is organized as follows. In Sec. II, we introduce the system, give the time evolutions of the expectation values of the system operators, and solve them. In Sec. III, the expressions for the components of the output field at the probe frequency and the Stokes frequency are given. In Sec. IV, we present the numerical results for the output probe field without or with the coupling field, and compare it with that from a ring cavity with two movable mirrors having equal frequencies. In Sec. V, we show the numerical results for the output Stokes field from a ring cavity with two moving mirrors having different or equal frequencies, and compare them. Finally in Sec. VI, we conclude the paper.

\section{Model}
\begin{figure}[!h]
\begin{center}
\scalebox{0.75}{\includegraphics{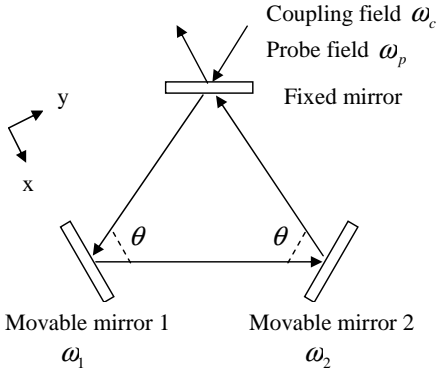}} \caption{\label{Fig1} The scheme of the optomechanical system. A strong coupling field at frequency $\omega_{c}$ and a weak probe field at frequency $\omega_{p}$ are injected into the ring cavity via the fixed mirror and interact with two movable mirrors whose resonance frequencies ($\omega_{1}, \omega_{2}$) are a little different.}
\end{center}
\end{figure}

 We consider a ring cavity with round-trip length $L$ formed by three mirrors, as shown in Fig. \ref{Fig1} \cite{Sumei}. One of them is not movable and partially transmitting, while the other two are allowed to vibrate and assumed to have 100\% reflectivity. The cavity field at the resonance frequency $\omega_{0}$ is driven by a strong coupling field with amplitude $\varepsilon$ at frequency $\omega_{c}$. Meanwhile a weak probe field with amplitude $\varepsilon_{p}$ at frequency $\omega_{p}$ is sent into the cavity. The coupling field and the probe field are treated classically here. The mechanical motions of both movable mirrors are coupled to the cavity field through the radiation pressure force exerted by the photons in the cavity. The movable mirrors' dynamics can be approximated as those of a single harmonic oscillator, with resonance frequency $\omega_{j}$, effective mass $m_{j}$, damping rate $\gamma_{j}$ ($j=1,2$).

 We assume that the system is in the adiabatic limit where the mechanical frequencies $\omega_{j}$ ($j=1,2$) are much smaller than the cavity free spectral range $c/L$ ($c$ is the speed of light in vacuum). The adiabatic limit $\omega_{j}\ll c/L$ ($j=1,2$) implies that the mechanical frequencies $\omega_{j}$ ($j=1,2$) are very small compared to the cavity resonance frequency $\omega_{0}$, so the moving mirrors are moving so slow that the retardation effect, the Casimir effect, and the Doppler effect become completely negligible \cite{Vitali63,Mancini,Vitali80}. Hence the radiation pressure force does not depend on the velocity of the movable mirrors. Assuming that $k$ is the wave vector of the cavity field with $k=\omega_{0}/c$, the radiation pressure force exerted by the light field on the movable mirror can be calculated from the momentum exchange between the cavity field and the movable mirror, which is $F=\frac{2\hbar k\cos(\theta/2)}{L/c}c^{\dag}c=\hbar\frac{2\omega_{0}}{L}c^{\dag}c\cos(\theta/2)$, where $\theta$ is the angle between the incident light and the reflected light at the surfaces of the movable mirrors, $c^{\dag}c$ is the photon number operator of the cavity field, $c^{\dag}$ and $c$ are the creation and annihilation operators of the cavity field, obey the standard commutation relation $[c,c^{\dag}]=1$. Note that the radiation pressure forces acting on the movable mirrors vary linearly with the instantaneous photon number in the cavity. Under the action of the radiation pressure force, the movable mirrors make small oscillations. Then the small motion of the mirrors changes the length of the optical cavity, and alters the intensity of the cavity field, which in turn modifies the radiation pressure force acting on the mirrors. Therefore the interaction between the cavity field and the mechanical motion in the ring cavity is nonlinear.

 In a frame rotating at the driving frequency $\omega_{c}$, the Hamiltonian of the whole system takes the form
\begin{eqnarray}\label{1}
H&=&\hbar(\omega_{0}-\omega_{c})c^{\dag}c+\frac{\hbar\omega_{1}}{2}(Q_{1}^2+P_{1}^{2})+\frac{\hbar\omega_{2}}{2}(Q_{2}^2+P_{2}^{2})\nonumber\\
& &+\hbar (g_{1}Q_{1}-g_{2}Q_{2})c^{\dag}c\cos\frac{\theta}{2}+i\hbar \varepsilon (c^{\dag}-c)\nonumber\\& &+i\hbar(\varepsilon_{p}c^{\dag}e^{-i\delta t}-\varepsilon^{*}_{p}ce^{i\delta t}).
\end{eqnarray}
 Here the first three terms are the free energies of the cavity field and two mechanical oscillators, respectively, $(Q_{j}, P_{j})$ denote the dimensionless position and momentum quadratures of the two mirrors, $Q_{j}=\sqrt{\frac{m_{j}\omega_{j}}{\hbar}}q_{j}$, $P_{j}=\sqrt{\frac{1}{m_{j}\hbar\omega_{j}}}p_{j}$ ($j=1, 2$), and $[Q_{j}, P_{k}]=i\delta_{j,k}$ ($j,k=1,2$). The fourth term describes the nonlinear optomechanical interactions between the cavity field and the two movable mirrors, $g_{j}=\frac{2\omega_{0}}{L}\sqrt{\frac{\hbar}{m_{j}\omega_{j}}}$ ($j=1, 2$) is the optomechanical coupling strength. The last two terms give the interactions of the cavity field with the coupling field and the probe field, respectively, $\varepsilon$ is related to the power $\wp$ of the coupling field by $\varepsilon=\sqrt{\frac{2\kappa \wp}{\hbar \omega_{c}}}$, where $\kappa$ is the decay rate of the cavity due to the transmission losses through the fixed mirror, $\varepsilon_{p}$ is related to the power $\wp_{p}$ of the probe field by $|\varepsilon_{p}|=\sqrt{\frac{2\kappa \wp_{p}}{\hbar \omega_{p}}}$, $\delta=\omega_{p}-\omega_{c}$ is the detuning of the probe field from the coupling field.

  Starting from the Heisenberg equations of motion, taking into account the dissipations of the cavity field and the mechanical oscillators, and neglecting quantum noise and thermal noise, we obtain the time evolutions of the expectation values of the system operators
\begin{eqnarray}\label{2}
\langle \dot{Q}_{1}\rangle&=&\omega_{1}\langle P_{1}\rangle,\nonumber\\
\langle \dot{P}_{1}\rangle&=&-\omega_{1}\langle Q_{1}\rangle-g_{1}\langle c^{\dag}\rangle \langle c\rangle\cos\frac{\theta}{2}-\gamma_{1}\langle P_{1}\rangle,\nonumber\\
\langle \dot{Q}_{2}\rangle&=&\omega_{2}\langle P_{2}\rangle,\nonumber\\
\langle \dot{P}_{2}\rangle&=&-\omega_{2}\langle Q_{2}\rangle+g_{2}\langle c^{\dag}\rangle \langle c\rangle\cos\frac{\theta}{2}-\gamma_{2}\langle P_{2}\rangle,\nonumber\\
\langle \dot{c}\rangle&=&-i[\omega_{0}-\omega_{c}+(g_{1}\langle Q_{1}\rangle-g_{2}\langle Q_{2}\rangle)\cos\frac{\theta}{2}]\langle c\rangle+\varepsilon\nonumber\\
& &+\varepsilon_{p}e^{-i\delta t}-\kappa \langle c\rangle,
\end{eqnarray}
where we have used the mean field assumption $\langle c^{\dag}c\rangle\simeq\langle c^{\dag}\rangle\langle c\rangle$ and $\langle Q_{j}c\rangle\simeq\langle Q_{j}\rangle\langle c\rangle$ ($j=1,2$). Since the probe field is much weaker than the coupling field $(|\varepsilon_{p}|<<\varepsilon)$,
the steady-state solution to Eq. (\ref{2}) can be approximated to the first order in the probe field $\varepsilon_{p}$. In the long time limit, the solution to Eq. (\ref{2}) can be written as
\begin{eqnarray}\label{3}
\langle s\rangle=s_{0}+s_{+}\varepsilon_{p}e^{-i\delta t}+s_{-}\varepsilon_{p}^{*}e^{i\delta t},
\end{eqnarray}
where $s=Q_{1}$, $P_{1}$, $Q_{2}$, $P_{2}$, or $c$. The solution contains three components, which in the original
frame oscillate at $\omega_{c}$, $\omega_{p}$, $2\omega_{c}-\omega_{p}$, respectively. Substituting Eq. (\ref{3}) into Eq. (\ref{2}), equating coefficients of $e^{0}$ and $e^{\pm i\delta t}$, we find the following analytical expressions
\begin{eqnarray}\label{4}
Q_{10}&=&-\frac{G_{1}}{\omega_{1}}|c_{0}|,\quad P_{10}=0,\nonumber\\
Q_{20}&=&\frac{G_{2}}{\omega_{2}}|c_{0}|,\quad P_{20}=0,\nonumber\\
c_{0}&=&\frac{\varepsilon}{\kappa+i\Delta'},\nonumber\\
c_{+}&=&\frac{1}{d(\delta)}\big\{[\kappa-i(\Delta'+\delta)](\omega_{1}^2-\delta^2-i\gamma_{1}\delta)\nonumber\\
& &\times(\omega_{2}^2-\delta^2-i\gamma_{2}\delta)+i[G_{1}^2\omega_{1}(\omega_{2}^2-\delta^2-i\gamma_{2}\delta)\nonumber\\
& &+G_{2}^2\omega_{2}(\omega_{1}^2-\delta^2-i\gamma_{1}\delta)]\big\},\nonumber\\
c_{-}&=&\frac{ic_{0}^2}{|c_{0}|^2d(\delta)^{*}}[G_{1}^2\omega_{1}(\omega_{2}^2-\delta^2+i\gamma_{2}\delta)\nonumber\\
& &+G_{2}^2\omega_{2}(\omega_{1}^2-\delta^2+i\gamma_{1}\delta)],
\end{eqnarray}
where $G_{1}=g_{1}|c_{0}|\cos\frac{\theta}{2}$ and  $G_{2}=g_{2}|c_{0}|\cos\frac{\theta}{2}$ are the effective optomechanical coupling rates,  $\Delta'=\omega_{0}-\omega_{c}+(g_{1}Q_{10}-g_{2}Q_{20})\cos\frac{\theta}{2}$ is the effective detuning of the coupling field from the cavity resonance frequency, including the frequency shift induced by the radiation pressure, and
\begin{eqnarray}\label{5}
d(\delta)&=&[\kappa+i(\Delta'-\delta)][\kappa-i(\Delta'+\delta)](\omega_{1}^2-\delta^2-i\gamma_{1}\delta)\nonumber\\& &\times(\omega_{2}^2-\delta^2-i\gamma_{2}\delta)-2\Delta'[G_{1}^2\omega_{1}(\omega_{2}^2-\delta^2-i\gamma_{2}\delta)\nonumber\\
& &+G_{2}^2\omega_{2}(\omega_{1}^2-\delta^2-i\gamma_{1}\delta)].
\end{eqnarray}

\section{The output field}

The output field can be obtained by using the input-output relation $\varepsilon_{out}(t)=2\kappa \langle c\rangle$.
In analogy with Eq. (\ref{3}), we expand the output field to the first order in the probe field $\varepsilon_{p}$,
\begin{eqnarray}\label{6}
\varepsilon_{out}(t)=\varepsilon_{out0}+\varepsilon_{out+}\varepsilon_{p}e^{-i\delta t}+\varepsilon_{out-}\varepsilon_{p}^{*}e^{i\delta t},
\end{eqnarray}
where $\varepsilon_{out0}$, $\varepsilon_{out+}$, and $\varepsilon_{out-}$ are the components of the output field oscillating at frequencies $\omega_{c}$, $\omega_{p}$, $2\omega_{c}-\omega_{p}$. Here $\varepsilon_{out-}$ is called a Stokes field, and it is generated via the nonlinear four-wave mixing process, in which two photons at frequency $\omega_{c}$ interact with a single photon at frequency $\omega_{p}$ to create a new photon at frequency $2\omega_{c}-\omega_{p}$. Thus we find that the components of the output field at the probe frequency and the Stokes frequency are
\begin{eqnarray}\label{7}
\varepsilon_{out+}&=&2\kappa c_{+},\nonumber\\
\varepsilon_{out-}&=&2\kappa c_{-},
\end{eqnarray}
respectively.
In the absence of the coupling field $(\wp=0)$, the components of the output field at the probe frequency and the Stokes frequency
are given by
\begin{eqnarray}\label{8}
\varepsilon_{out+}&=&\frac{2\kappa}{\kappa+i(\Delta'-\delta)},\nonumber\\
\varepsilon_{out-}&=&0.
\end{eqnarray}
These are not unexpected results.
Let us write the real part of $\varepsilon_{out+}$ as $\nu_{p}$, which exhibits the absorption
characteristic of the output field at the probe frequency. It can be measured by the homodyne technique \cite{Walls}.

\section{Numerical results of the output probe field}
In this section, we numerically evaluate how the coupling field modifies the absorption of the ring cavity to the probe field.

\begin{figure}[!h]
\begin{center}
\scalebox{0.65}{\includegraphics{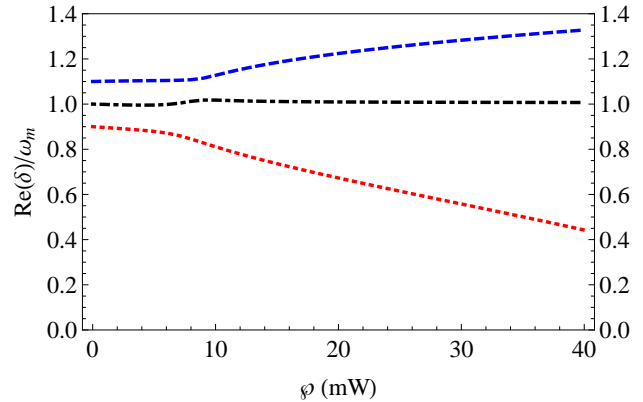}} \caption{\label{Fig2} (Color online) The dependence of the real parts of the roots of $d(\delta)$ in the domain $\mbox{Re}(\delta)>0$ on the power $\wp$ of the coupling field.}
\end{center}
\end{figure}
\begin{figure}[!h]
\begin{center}
\scalebox{0.65}{\includegraphics{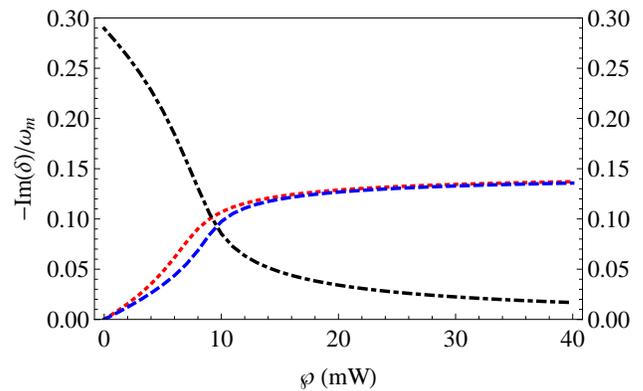}} \caption{\label{Fig3} (Color online) The dependence of the imaginary parts of the roots of $d(\delta)$ on the power $\wp$ of the coupling field.}
\end{center}
\end{figure}
The values of the parameters chosen are similar to those in \cite{Weis}: the wavelength of the coupling field $\lambda=2\pi c/\omega_{c}=775$ nm, the coupling constants $g_{1}=2\pi\times 12$ GHz/nm $\times\sqrt{\hbar/(m_{1}\omega_{1})}$, $g_{2}=2\pi\times 12$ GHz/nm $\times\sqrt{\hbar/(m_{2}\omega_{2})}$, the masses of the movable mirrors  $m_{1}=m_{2}=20$ ng, the frequencies of the movable mirrors $\omega_{1}=\omega_{m}+0.1 \omega_{m}$, $\omega_{2}=\omega_{m}-0.1 \omega_{m}$, where $\omega_{m}=2\pi\times 51.8$ MHz, the cavity decay rate $\kappa=2\pi\times15$ MHz, $\kappa/\omega_{m}\simeq0.289<1$ (the system is placed in resolved sideband regime), the mechanical damping rates $\gamma_{1}=\gamma_{2}=\gamma=2\pi\times 4.1$ kHz, the mechanical quality factors $Q'_{1}=\omega_{1}/\gamma_{1}\simeq13897$, $Q'_{2}=\omega_{2}/\gamma_{2}\simeq11370$, the angle $\theta=\pi/3$. And the coupling field is tuned close to the red sideband of the cavity resonance $\Delta'=\omega_{m}$. The parameters chosen ensure the system operating in the stable regime.

We note that the structure of the quadrature of the output probe field $\nu_{p}$ is determined by $d(\delta)$. The roots $\delta$ of $d(\delta)$ are complex values. The real parts $\mbox{Re}(\delta)$ of the roots determine the
positions of the normal modes of the optomechanical system; the imaginary parts $\mbox{Im}(\delta)$ of the roots describe their widths. Figure \ref{Fig2} shows the real parts of the roots of $d(\delta)$ in the domain $\mbox{Re}(\delta)>0$ versus the coupling beam power. It is seen that the real parts of the roots of $d(\delta)$ are $\mbox{Re}(\delta)=0.9\omega_{m}, \omega_{m}, 1.1\omega_{m}$ at low driving power. At high driving power, one of the real parts is still $\omega_{m}$ (dotdashed curve), not changing with increasing the power of the coupling field, the difference between the other two (dotted curve and dashed curve) increases with increasing the power of the coupling field, which implies the normal mode spitting \cite{Florian,Kippenberg,Aspelmeyer} in the output probe field. Figure \ref{Fig3} shows the imaginary parts of the roots of $d(\delta)$ versus the coupling beam power. We see that the imaginary parts of the roots of $d(\delta)$ have three different values at low driving power. At high driving power, the imaginary parts of the roots of $d(\delta)$ have three values, two of them are identical, the other one is small. Moreover, in Figs. \ref{Fig2} and \ref{Fig3}, the dotdashed curves represent the real part and the imaginary part of one of the roots of $d(\delta)$, respectively, similarly for dotted curves and dashed curves. Hence, at high driving power, the central peak in the quadrature $\nu_{p}$ is narrow, two side peaks in the quadrature $\nu_{p}$ have the same broad linewidths.

\begin{figure}[!h]
\begin{center}
\scalebox{0.65}{\includegraphics{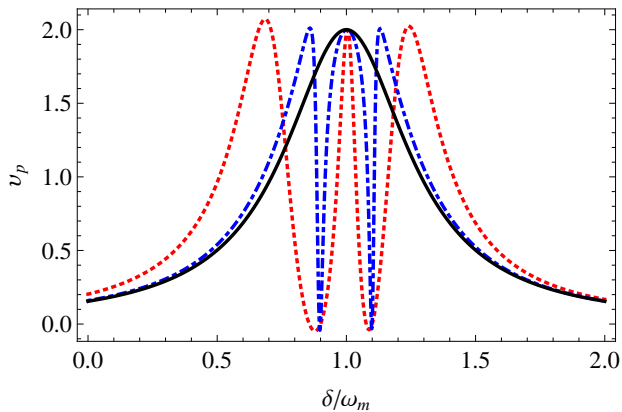}} \caption{\label{Fig4} (Color online) The quadrature of the output probe field $\nu_{p}$ as a function of the normalized probe detuning $\delta/\omega_{m}$ for $\omega_{1}=1.1\omega_{m}$ and $\omega_{2}=0.9\omega_{m}$. The black solid, blue dotdashed, and red dotted curves correspond to $\wp=0$, 2 mW, and 15 mW, respectively.}
\end{center}
\end{figure}
In Fig. \ref{Fig4}, we plot the quadrature of the output probe field $\nu_{p}$ as a function of the normalized probe detuning $\delta/\omega_{m}$ for three different powers of the coupling field. In the absence of the coupling field, $\nu_{p}$ (solid curve) has a standard Lorentzian absorption peak with a full width at half maximum (FWHM) of $2\kappa$ at the line center $\delta/\omega_{m}=1$. However, in the presence of the coupling field with power 2 mW, it is seen that the dotdashed curve exhibits two symmetric narrow EIT dips centered at $\delta/\omega_{m}=0.9,1.1$ and a broad absorption peak centered at $\delta/\omega_{m}=1$. The FWHM of the two EIT dips are about $(\gamma+\frac{G_{1}^2}{\kappa})$ and $(\gamma+\frac{G_{2}^2}{\kappa})$, respectively. The FWHM of the central absorption peak is about $(\omega_{1}-\omega_{2})-(\gamma+\frac{G_{1}^2+G_{2}^2}{2\kappa})$. The two transparency dips display that the input probe field could be simultaneously transparent at two symmetric frequencies, which is the result of the destructive interferences between the probe field and the anti-Stokes fields at frequencies $\omega_{c}+0.9\omega_{m}$ and $\omega_{c}+1.1\omega_{m}$ generated by the interactions of the coupling field with the movable mirrors. The absorption peak at the line center implies that the incident probe field is almost fully absorbed by the optomechanical system. Moreover, increasing the power $\wp$ of the coupling field, the two EIT dips become broader, which results in narrowing of the central absorption peak. Further, from the dotted curve in Fig. \ref{Fig4}, one can see that increasing the power of the coupling field to $15$ mW results in a larger splitting between the left and right side peaks with the same linewidths, it also results in a narrow central peak. Our calculations show that in the strong coupling limit $2(G_{1}^2+G_{2}^2)>>(\kappa-\frac{\gamma}{2})^2$, the three dressed modes of the system corresponding to the three absorption peaks are $\delta\approx\omega_{m}-i\gamma/2$, and $\delta\approx\omega_{m}\pm\frac{1}{2}\sqrt{2(G_{1}^2+G_{2}^2)}-\frac{i}{2}(\kappa+\frac{\gamma}{2})$.  Note that the FWHM for the narrow central peak at $\delta=\omega_{m}$ is about $\gamma$. In addition, the FWHM for the right and left peaks at $\delta\approx\omega_{m}\pm\frac{1}{2}\sqrt{2(G_{1}^2+G_{2}^2)}$ are the same $(\kappa+\frac{\gamma}{2})$, the splitting between the side peaks is about $\sqrt{2(G_{1}^2+G_{2}^2)}$, which is proportional to the power of the coupling field. These results are consistent with those in Figs. 2 and 3.

\begin{figure}[!h]
\begin{center}
\scalebox{0.65}{\includegraphics{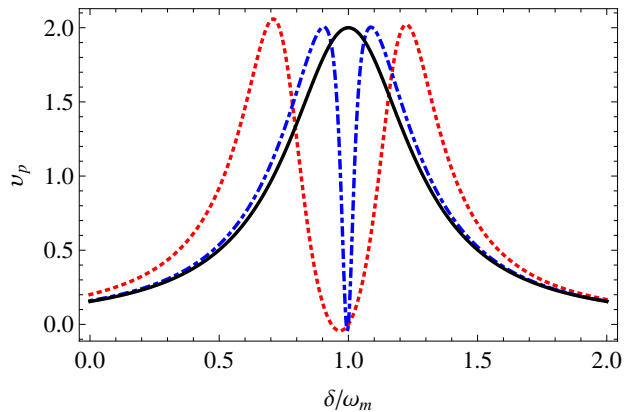}} \caption{\label{Fig5} (Color online) The quadrature of the output probe field $\nu_{p}$ as a function of the normalized probe detuning $\delta/\omega_{m}$ for $\omega_{1}=\omega_{2}=\omega_{m}$. The black solid, blue dotdashed, and red dotted curves correspond to $\wp=0$, 2 mW, and 15 mW, respectively.}
\end{center}
\end{figure}

For comparison, we consider the same previous system, but the two mechanical oscillators have the same frequencies $\omega_{1}=\omega_{2}=\omega_{m}$, thus their effective optomechanical coupling rates are equal, we assume $G_{1}=G_{2}=G$. We plot the quadrature of the output probe field $\nu_{p}$ as a function of the normalized probe detuning $\delta/\omega_{m}$ for three different powers of the coupling field, as shown in Fig. \ref{Fig5}. When the coupling field is not present, the quadrature $\nu_{p}$ (solid curve) has a Lorentzian absorption lineshape. However, the presence of the control field with power 2 mW leads to a narrow EIT-like dip at the line center (dotdashed curve). Thus the probe field can almost completely propagate through the ring cavity on resonance with almost no absorption. The FWHM of the EIT-like dip is about $\gamma+\frac{2G^2}{\kappa}$. The EIT-like dip is attributed to the destructive interference between the input weak probe field and the scattering quantum fields at the probe frequency $\omega_{p}$ generated by the interactions of the coupling field with two mirrors having identical frequencies. Moreover, in the strong coupling limit $2G>>\kappa$, the normal mode splitting exhibits in the quadrature $\nu_{p}$ (dotted curve), the two peaks have the same FWHM of $\kappa+\frac{\gamma}{2}$, their positions are $\delta\approx\omega_{m}\pm G$, the separation between them is about $2G$.

In order to understand the narrow central peak in Fig. \ref{Fig4}  and the EIT-like dip in Fig. \ref{Fig5}, let us introduce the relative coordinates ($Q_{a}$, $P_{a}$) and center of mass coordinates ($Q_{s}$, $P_{s}$) of the two movable mirrors as
\begin{eqnarray}
Q_{a}&=&\frac{g_{1}Q_{1}-g_{2}Q_{2}}{\sqrt{g_{1}^2+g_{2}^2}},\quad P_{a}=\frac{g_{1}P_{1}-g_{2}P_{2}}{\sqrt{g_{1}^2+g_{2}^2}},\nonumber\\
Q_{s}&=&\frac{g_{1}Q_{1}+g_{2}Q_{2}}{\sqrt{g_{1}^2+g_{2}^2}},\quad P_{s}=\frac{g_{1}P_{1}+g_{2}P_{2}}{\sqrt{g_{1}^2+g_{2}^2}}.
\end{eqnarray}
With these new coordinates we can write the Hamiltonian Eq. (\ref{1}) as
\begin{eqnarray}\label{10}
H&=&\hbar(\omega_{0}-\omega_{c})c^{\dag}c+\frac{\hbar\omega}{2}(Q_{a}^2+P_{a}^{2})+\frac{\hbar\omega}{2}(Q_{s}^2+P_{s}^{2})\nonumber\\
& &+\frac{\hbar}{4}(g_{1}^2+g_{2}^2)(\frac{\omega_{1}}{g_{1}^2}-\frac{\omega_{2}}{g_{2}^2})(Q_{a}Q_{s}+P_{a}P_{s})\nonumber\\
& &+\hbar \sqrt{g_{1}^2+g_{2}^2}Q_{a}c^{\dag}c\cos\frac{\theta}{2}+i\hbar \varepsilon (c^{\dag}-c)\nonumber\\& &+i\hbar(\varepsilon_{p}c^{\dag}e^{-i\delta t}-\varepsilon^{*}_{p}ce^{i\delta t}),
\end{eqnarray}
where $\omega=\frac{1}{4}(g_{1}^2+g_{2}^2)(\frac{\omega_{1}}{g_{1}^2}+\frac{\omega_{2}}{g_{2}^2})$. The fourth term in Eq. (\ref{10}) describes the interaction between the relative and center of mass coordinates. The fifth term in Eq. (\ref{10}) shows that the interaction between the relative coordinate and the cavity field. When the two mechanical frequencies are equal $\omega_{1}=\omega_{2}$, the fourth term vanishes, the center of mass coordinate is decoupled from the relative coordinate, and it is also decoupled from the cavity field, so only the relative coordinate is coupled to the cavity field via radiation pressure, which generates a transparency dip in the output probe field, as shown in Fig. \ref{Fig5}. When the two mechanical frequencies are not equal $\omega_{1}\neq\omega_{2}$, it is seen that not only the relative coordinate and the cavity field interact with each other, but also the relative and center of mass coordinates interact with each other, which leads to a narrow central peak in the output probe field, as shown in Fig. \ref{Fig4}.

\section{Numerical results of the output Stokes field}

 In this section, we numerically examine the effect of the coupling field on the output Stokes field generated via the four-wave mixing process.

\begin{figure}[!h]
\begin{center}
\scalebox{0.65}{\includegraphics{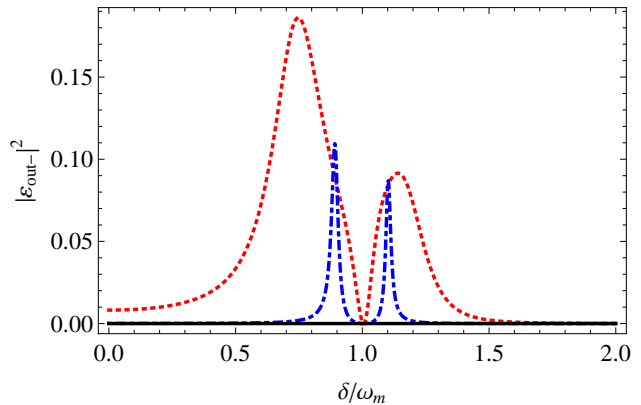}} \caption{\label{Fig6} (Color online) The intensity of the Stokes field $|\varepsilon_{out-}|^2$ as a function of the normalized probe detuning $\delta/\omega_{m}$ for $\omega_{1}=1.1\omega_{m}$ and $\omega_{2}=0.9\omega_{m}$. The black solid, blue dotdashed, and red dotted curves correspond to $\wp=0$, 2 mW, and 15 mW, respectively.}
\end{center}
\end{figure}

Figure \ref{Fig6} shows the intensity of the generated Stokes field $|\varepsilon_{out-}|^2$ as a function of the normalized probe detuning $\delta/\omega_{m}$ for several values of the driving power when the frequencies of the two mirrors are not equal ($\omega_{1}=1.1\omega_{m}$ and $\omega_{2}=0.9\omega_{m}$). It is seen that $|\varepsilon_{out-}|^2=0$ (solid curve) in the absence of the coupling field. However, in the presence of the coupling field, the $|\varepsilon_{out-}|^2$ (dotdashed curve and dotted curve) exhibits the normal mode splitting, the peak separation increases with the power of the coupling field. Moreover, the maximum value of $|\varepsilon_{out-}|^2$ is increased with the power of the coupling field, the maximum value of $|\varepsilon_{out-}|^2$ is about 0.19 when $\wp=15$ mW. Note that the intensity of the Stokes field goes to zero when the detuning $\delta/\omega_{m}=1$. Hence, the four-wave mixing effect is completely suppressed on resonance so that there is only the probe field stored inside the optomechanical system. The suppression of the four-wave mixing effect on resonance is due to the destructive interference between the Stokes fields produced by the coupling field interacting with the two movable mirrors oscillating at frequencies $\omega_{1}=1.1\omega_{m}$ and $\omega_{2}=0.9\omega_{m}$.

\begin{figure}[!h]
\begin{center}
\scalebox{0.65}{\includegraphics{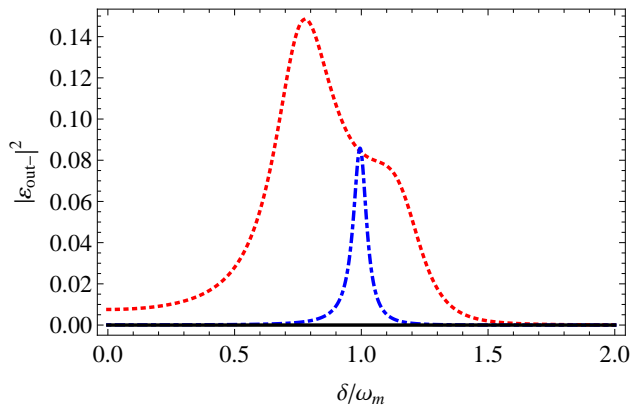}} \caption{\label{Fig7} (Color online) The intensity of the Stokes field $|\varepsilon_{out-}|^2$ as a function of the normalized probe detuning $\delta/\omega_{m}$ for $\omega_{1}=\omega_{2}=\omega_{m}$. The black solid, blue dotdashed, and red dotted curves correspond to $\wp=0$, 2 mW, and 15 mW, respectively.}
\end{center}
\end{figure}

For comparison, we also consider the case of a ring cavity in which the frequencies of the two mirrors are equal ($\omega_{1}=\omega_{2}=\omega_{m}$), the intensity $|\varepsilon_{out-}|^2$ of the output Stokes field versus the normalized probe detuning $\delta/\omega_{m}$ for several values of the driving power is shown in Fig. \ref{Fig7}. The significant difference between Fig. \ref{Fig7} and Fig. \ref{Fig6} is that the intensity of the Stokes field in Fig. \ref{Fig7} is non zero at $\delta/\omega_{m}=1$ in the presence of the coupling field. Thus the four-wave mixing process is not suppressed at $\delta/\omega_{m}=1$, which arises from the constructive interference between the Stokes fields produced by the coupling field interacting with the two movable mirrors oscillating at the same frequencies $\omega_{1}=\omega_{2}=\omega_{m}$. So there is no apparent normal mode splitting in this case.

\section{Conclusions}
 To summarize, we have demonstrated how a strong coupling field affects the propagation of a weak probe field in a ring cavity with two movable mirrors whose frequencies are close to each other. We find double EIT dips and a central absorption peak in the output probe field in the presence of the coupling field. Thus this system can become transparent at two different frequencies of a weak probe field. Increasing the pump power gives rise to broadening two EIT dips and narrowing the central absorption peak. In addition, we show that the coupling field induces the normal mode splitting in the output probe field and the output Stokes field. Moreover, the four-wave mixing suppression takes place in this optomechanical system when the probe detuning is one-half the sum of the two mechanical frequencies. However, when two movable mirrors in a ring cavity have identical frequencies, the response of the ring cavity to a weak probe field in the presence of a strong coupling field would be different. We observe a narrow transparency dip in the output probe field, and no four-wave mixing suppression when the probe detuning is equal to the mechanical frequency.

The author thanks Prof. G. S. Agarwal for helpful discussions. The author also thanks Prof. Lin Tian and Prof. Mankei Tsang for support.

\end{document}